\documentclass[prb,onecolumn,nobibnotes,groupedaddress]{revtex4}

\usepackage{natbib}
\usepackage{algorithm,algorithmic}
\usepackage{enumerate}
\usepackage{listings}
\usepackage{color}
\usepackage{float}
\usepackage{amssymb}
\usepackage{amsmath}
\usepackage{blochsphere}
\usepackage{tikz}
\usetikzlibrary{quantikz}

\usepackage{nicematrix}

\usepackage{graphicx}

\begin{document}

\title{The Basics of Quantum Computing for Chemists}
\author{Daniel Claudino$^{1,2,3*}$}
\affiliation{$^{1}$ Quantum Computing Institute,\ Oak\ Ridge\ National\ Laboratory,\ Oak\ Ridge,\ TN,\ 37831,\ USA \\
$^{2}$Computational Sciences and Engineering Division,\ Oak\ Ridge\ National\ Laboratory,\ Oak\ Ridge,\ TN,\ 37831,\ USA \\
$^{3}$Computer Science and Mathematics Division,\ Oak\ Ridge\ National\ Laboratory,\ Oak\ Ridge,\ TN,\ 37831,\ USA}
\email[]{claudinodc@ornl.gov}
\thanks{\\This manuscript has been authored by UT-Battelle, LLC, under Contract No.~DE-AC0500OR22725 with the U.S.~Department of Energy. The United States Government retains and the publisher, by accepting the article for publication, acknowledges that the United States Government retains a non-exclusive, paid-up, irrevocable, world-wide license to publish or reproduce the published form of this manuscript, or allow others to do so, for the United States Government purposes. The Department of Energy will provide public access to these results of federally sponsored research in accordance with the DOE Public Access Plan.}

\begin{abstract}

The rapid and successful strides in quantum chemistry in the past decades can be largely credited to a conspicuous synergy between theoretical and computational advancements. However, the architectural computer archetype that enabled such a progress is approaching a state of more stagnant development. One of the most promising technological avenues for the continuing progress of quantum chemistry is the emerging quantum computing paradigm. This revolutionary proposal comes with several challenges, which span a wide array of disciplines. In chemistry, it implies, among other things, a need to reformulate some of its long established cornerstones in order to adjust to the operational demands and constraints of quantum computers. Due to its relatively recent emergence, much of quantum computing may still seem fairly nebulous and largely unknown to most chemists. It is in this context that here we review and illustrate the basic aspects of quantum information and their relation to quantum computing insofar as enabling simulations of quantum chemistry. We consider some of the most relevant developments in light of these aspects and discuss the current landscape when of relevance to quantum chemical simulations in quantum computers.

\end{abstract}


\maketitle
\section{Introduction}

One of the main interests shared by the chemistry community at large can be summarized by the title of the presentation delivered by Nobel Prize laureate Robert S. Mulliken upon his acceptance of the G.N. Lewis Award: ``What are the Electrons Really Doing in Molecules?''.\cite{Adams00roberts.} From a quantum chemistry perspective, the early recognition that the electrons make up the bulk of the quantum effects in molecules led to the pervasiveness of the Born-Oppenheimer approximation\cite{boa} in chemistry. It has enabled quantum chemistry to provide decisive answers to many questions, and albeit the great deal of simplification provided by the abovementioned approximation, the underlying electronic problem remains a perennial challenge and such a status will likely stand for the foreseeable future. One of the sources of this predicament rises from the necessity to resort to expressing a many-body wave function in terms of a one-particle basis, more commonly referred to as spin-orbitals. Looking at the electronic structure problem through the second quantization formalism,\cite{jorgensen, shavitt_bartlett_2009} in which quantum states are identified by the occupation of these spin-orbitals, it becomes clear that associated Hilbert space grows exponentially with the number of orbitals $N$ in a given physical system. Taking advantage of symmetries, such as conservation of particle number and spin, not only provides a way to enforce physically important restrictions but also adds the benefit of constraining the search for the solution to just a subspace of the 2$^N$-dimensional Hilbert space (a sector of Fock space when speaking of particle number conservation). Yet, it still consists of a combinatorial problem that scales as $N\choose k$, for it is akin to arranging $k$ electrons in $N$ spin-orbitals, hence exhibiting an undesirable factorial growth.

Despite these hurdles, quantum chemistry has experienced tremendous success in the past few decades, and much of this ascension can be credited to a synergy of advancements in three main fronts. The first is the introduction of several formally and physically sound approximate methods which circumvent the overwhelming dimensionality restriction while delivering reliable answers. Because much of quantum chemistry lacks a closed-form solution, which is true of quantum mechanics in general, it relies on complex numerical techniques whose realization became possible with the emergence of the digital electronic computer. This prompted the development of specialized software, mostly in the form of efficient libraries of mathematical routines to carry out fundamental operations present in the algorithms that are employed to compute quantities of quantum chemical interest, such as efficient linear algebra protocols and optimized numerical integration packages.\cite{integrals} The continuous evolution of the hardware capabilities of these computers, such as the manufacture of better and smaller transistors and an ever-growing number of processors capable of functioning in concerted operation, culminated in the concept of high-performance computing (HPC) clusters. Thus, the confluence of these three aspects, namely theory, hardware, and software helped push the boundaries and extended the size and time scale of what was deemed possible in quantum chemistry. 

For some time, the underlying trend in the observed improvement of hardware capabilities was remarkable and became known as the Moore's Law.\cite{moore1, moore2} However, the initially observed pattern of roughly doubling the density of transistors in a chip every 18-24 months is showing signs of slowing down, calling for new computing paradigms to meet the growing computational demands in several scientific and technological fields.\cite{moore3} Among the candidates expected to provide the next \textit{quantum} leap in computational technology, quantum computing emerges as the most promising. Yet, because quantum computing is still in its infancy, it is a long way towards capabilities exceeding what has been achieved through classical compute machinery. Tying back to the main topic of this manuscript, this means that once more a great effort spanning many different disciplines is expected to be necessary in order to elevate quantum computers to be the main workhorse behind the quantum chemistry inquiry.

The initial case for quantum computing came from Richard Feynman,\cite{Feynman1982} who is often credited for conceiving the idea that a computing device working under the laws of quantum mechanics could be exponentially faster than a classical machine for certain classes of problems, an idea which was independently reached somewhat earlier by Manin.\cite{manin} One of the most immediate and natural applications of a quantum computer envisioned by Feynman was the simulation of other quantum systems. On one hand, this opens the door to the solution of many problems that would otherwise be ``practically impossible.'' On the other hand, as it will become clear later, it often demands a complete rethinking of these problems, and quantum chemistry is no exception, which practically means revisiting the three foundational pillars that helped elevate quantum chemistry in the first place. The phrase ``practically impossible'' appears in quotes for a reason. For our purposes, it means that problems whose resource demands do not scale polynomially with the size/time of the simulation. In other words, if upon varying the dimension of the Hilbert space or the propagated time one observes an exponential response in the computational resources necessary to carry out the underlying computation, this signals that this type of simulation is very limited in scope. The question of which problems are practically possible in this sense is addressed in a rich field of theoretical computer science known as computational complexity theory. However, such considerations are beyond the practical concerns of most chemists and will not be addressed here.

To start putting things in perspective, one can look at quantum chemistry from the standpoint of electronic structure theory, which has long benefited from the second quantization formalism, representing states in an occupation basis, i.e., Fock space.\cite{jorgensen, shavitt_bartlett_2009} The exact eigenspectrum for a non-relativistic Hamiltonian in a given basis can be found by diagonalization of the full Hamiltonian, amounting to solving the full configuration interaction (FCI) problem. Yet, it is easy to verify that the dimension of the Hilbert space in question grows as 2$^N$, with $N$ being the number of basis functions, that is, exhibits exponential growth. In practical terms, the number of matrix elements that need to be evaluated scales as 4$^N$, alleviated to $2^N(2^N+1)/2$ in the case of an observable, such as the molecular Hamiltonian, due to the Hermiticity constraint. Turning to the dynamics problem, the textbook example of a separable wave function in terms of time and spatial domain points to an exponential dependence with the time variable, which holds in the adiabatic approximation (the Hamiltonian is constant or varies only slowly in time). Thus, in simple terms, such methods are amenable only for implementation within a compute paradigm where information can be stored and processed exponentially with the number of resources available. The great potential from quantum computing is due to the superposition and entanglement found among qubits, which enables a quantum computer to naturally achieve such an exponential increase in performance.

In light of Feynman's proposal and keeping in mind that chemistry is also governed by the laws of quantum mechanics, quantum chemistry has been seen by many as one of the first fields to showcase a tangible advantage from the employment of quantum computers. Yet, for a long time the majority of the quantum algorithms were not directly related to chemistry, such as the Bernstein–Vazirani algorithm to determine an unknown string of bits (bit string)\cite{bernstein_vazirani} and its more general Deutsch–Jozsa algorithm,\cite{deutsch_jozsa} Grover's search algorithm,\cite{grover} Shor's prime number factoring,\cite{shor} etc. The Quantum Phase Estimation (QPE) algorithm is pervasive in many other quantum computing algorithms, and even though it is not specific to quantum chemistry, it allows one to compute an eigenvalue for a given eigenstate of an observable with tunable precision. This made the QPE algorithm a prominent candidate in quantum chemical simulations, since many quantum chemistry methods aim at just that, particularly at recovering the ground state energy via the diagonalization of the Hamiltonian in a basis of trial wave functions that are variationally optimized. However, its application to even relatively small molecules/basis sets is accompanied by quantum circuits with prohibitive implementation at present.

Given the early stage of quantum computing at the moment, understanding the availability and demand of quantum resources is of paramount importance. In evaluating the suitability of a quantum algorithm, a crucial metric is the so-called \textit{circuit depth}. In short, the entire algorithm is comprised of a time-ordered series of relatively simple operations (gates) and each takes some amount of time to be executed. As we add more and more of these \emph{gates} (these will be better explained later), the more time it requires to complete the circuit that executes an algorithm. However, qubits can survive in a defined state for a limited time before external effects push them out of that state, negatively impacting the reliability of the information they can provide. The length of time in which assertions about the state can be made reliably is known as \textit{coherence time}. Because present qubits have shown significant resistance to endure for prolonged coherence times, it is one of the main limiting factors preventing quantum computers to reach their expected performance.  Another factor in curtailing a more widespread adoption of quantum computers is the small-to-moderate number of qubits in the current quantum chips, which are collectively known as noisy intermediate-scale quantum (NISQ) devices.\cite{Preskill2018quantumcomputingin} While these current technological challenges severely curtail the performance and operation of quantum computers, they have spurred the development of a new compute paradigm with the goal of circumventing the inherent shortcomings of the present technology. It recognizes that the quantum computer is better suited to specific tasks, such as the preparation of highly entangled states and the associated measurements, from which some cost function can be computed given a set of parameters that can tune the quantum circuit. The output of the cost function (a scalar) is provided to a classical optimization routine that in turn searches the parameter space in order to meet user-defined optimal criteria. It is this interplay between classical and quantum technologies that has led to the rise of hybrid quantum-classical computing. From a computer science/software engineering standpoint, this idea can be seen as a natural extension of heterogeneous compute approaches, highlighted by the concerted usage of central and graphical processing units (CPUs and GPUs, respectively), and has proven a pivotal move in the development of quantum computing at large as it constitutes an avenue for showcasing the strengths of quantum computers even before it reaches full maturity, a future milestone that is commonly referred to as the fault-tolerance regime.\cite{ftqc1, ftqc2}

The main goal of this review is to gently introduce quantum chemistry practitioners to the most important aspects of quantum computing that are of relevance in quantum chemistry. Basic understanding of quantum mechanics and its linear algebra formulation, tied back to Dirac's postulates,\cite{dirac2012the, shankar1994principles} is assumed as well as familiarity with quantum chemistry, primarily electronic structure theory and its second quantization formulation, as exposed in Szabo and Ostlund.\cite{szabo} The most prominent topics will be illustrated largely in light of quantum computing applications in the domain of electronic structure theory. The extension to problems of relevance in dynamics is believed to only be accessible in the fault-tolerant stage of quantum computing. Furthermore, this review is not intended to give a comprehensive overview of every single development in quantum computing that touches on quantum chemistry or vice-versa. For this purpose, the foundation given by this manuscript can be augmented by lengthier, more encompassing reviews.\cite{mcardle2020quantum, review1, review2} Valuable resources for the beginner researcher looking into a more practical take on quantum computing is the Qiskit textbook\cite{qiskit_book} and the tutorial in Ref.\citenum{beginner}. Finally, the quintessential textbook in the field of quantum computing is the one authored by Nielsen and Chuang.\cite{Nielsen2011}

\section{The basics of quantum computing}

\subsection{Information and bits}
\label{ssec:info}

In analogy with classical computing, where the bit is the basic building block of information, we have the quantum bit, yet with key differences from its classical counterpart. A classical bit, simply bit hereafter, is only ever found in two different ``states'', $|0\rangle$ and  $|1\rangle$, often associated with whether or not a transistor is charged, and consequently can be ascribed a binary value: 0 or 1. In principle, the state of the transistor and the value of the associated bit is always known, (or knowable), except when faults occur, which are relatively rare in current classical computers, but are still the focus of ongoing research.

\begin{figure}[ht!]
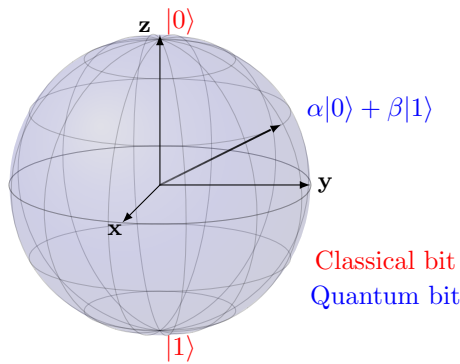

    \centering
    				\begin{blochsphere}[radius=2.0 cm,tilt=15,rotation=-20,opacity=0.1, color=blue]
                    \drawBallGrid[style={opacity=0.2}]{30}{30}
                    
                    \labelLatLon{up}{90}{0};
                    \labelLatLon{down}{-90}{90};
                    \node[above, anchor=south west] at (up) {\color{red}\fontsize{0.35cm}{1em}\selectfont $|0\rangle$};
                    \node[below, anchor=north west] at (down) {\color{red}\fontsize{0.35cm}{1em}\selectfont $|1\rangle$};
                    \drawStatePolar{qstate}{60}{0};
                    \draw [->] (0,0) -- (0,2);
                    \draw [->] (0,0) -- (2,0);
                    \draw [->] (0,0) -- (-0.5,-0.5);
                    \node at (2.2,0) {\textbf{y}};
                    \node at (-0.2, 2.1) {\textbf{z}};
                    \node at (-0.6,-0.6) {\textbf{x}};

                    \node at (2.8,1) {\color{blue}\fontsize{0.35cm}{1em}\selectfont $\alpha|0\rangle+\beta|1\rangle$};
                    \node at (3,-1) {\color{red}\fontsize{0.35cm}{1em}\selectfont Classical bit};
                    \node at (3,-1.5) {\color{blue}\fontsize{0.35cm}{1em}\selectfont Quantum bit};
                \end{blochsphere}
    \caption{\label{fig:bloch}The Bloch sphere.}
    
\end{figure}

On the other hand, a quantum bit, or a qubit for short, is in principle any two-level quantum system, such as the $\uparrow$ and $\downarrow$, or $\alpha$ and $\beta$ spin states. Realistically, the two states that characterize a qubit are rarely the only possible states, but two-level systems that are far removed from other states by large energy gaps and/or can only be achieved through transitions that undergo symmetry violation tend to be make for suitable qubits. Given the fact that qubits are bound by the laws of quantum mechanics, in principle any superposition of $|0\rangle$ and  $|1\rangle$ is possible. This is often illustrated by the Bloch's sphere, a pictorial representation of the states that can be assumed by a single qubit, shown in Figure \ref{fig:bloch}.

While the classical bit can only be found in the two poles of the sphere along the z-axis, the surface of the sphere is comprised of all superpositions of the two states. The ability of superimposing quantum states is partly responsible for the speedup that can be harnessed from quantum computers. In contrast to classical bits, which operate can always be ascribed individual states, two or more qubits can be found in \emph{entanglement}. Entanglement can be identified with qubits being found in a superposition of states that results in a uniquely quantum kind of correlation that prevents any qubit to be able to be addressed independently from the other qubits it is entangled with. And it is the entanglement property in conjunction with superposition, which are exclusive to quantum entities, the responsible factors for the attractive prospect speedup of quantum computing.

From the superposition property in a single qubit, we can write an arbitrary qubit state $|\psi \rangle$ as

\begin{equation}
    |\psi\rangle = \alpha|0\rangle+\beta|1\rangle = \begin{pmatrix} \alpha \\ \beta \end{pmatrix} = \alpha \begin{pmatrix} 1 \\ 0 \end{pmatrix} + \beta \begin{pmatrix} 0 \\ 1 \end{pmatrix},
\end{equation}
which exposes the fact that $|0\rangle = \begin{pmatrix} 1 \\ 0 \end{pmatrix}$ and $|1\rangle = \begin{pmatrix} 0 \\ 1 \end{pmatrix}$ can serve as a suitable basis to span the space of single qubit states. We can view altering the state of a qubit as performing a linear operation onto its vector representation. For the resulting state to remain in the same Hilbert space, or equivalently, to map onto an equal-dimensional Hilbert space, such mappings must be norm-preserving. Given that the qubit state is represented by two-dimensional vectors, we are looking to a set of 2$\times$ 2 unitary matrices, that is, the matrices that make up the $\mathfrak{su}(2)$ Lie algebra:

\begin{equation}
    \sigma_x = \begin{pmatrix} 0 & 1 \\ 1 & 0 \end{pmatrix}, \quad \sigma_y = \begin{pmatrix} 0 & -i \\ i & 0 \end{pmatrix}, \quad \sigma_z = \begin{pmatrix} 1 & 0 \\ 0 & -1 \end{pmatrix}
\end{equation}
These matrices, together with the identity $I_2$, can be used to form the generators of the operations onto qubit states and comprise the SU(2) Lie group.

While superposition can be observed in a single qubit, entanglement is a collective characteristic of quantum entities. This concept is closely related to the idea of correlated particles in quantum chemistry/many-body physics, which implies that a collection of correlated/entangled particles cannot be correctly treated independent of one another. At this stage we recall that the rationale for quantum computation put forth by Feynman was to simulate quantum entities. For the purposes of this exposition, it is worth noting that the correlated motions of electrons in matter is often a direct result of entanglement across many quantum degrees of freedom, a key feature that quantum computers are designed to provide. The reader might remember that the many-body wave function in independent particle theories, such as HF and DFT, can be written as a product of one-particle wave functions, which in turn do not account for electron correlation (not in a proper manner anyway). Conversely, the wave function of entangled particles cannot be simplified to a product form. The simplest and most common examples of entanglement are the Bell states:

\begin{align}
    \frac{1}{\sqrt{2}}(|00\rangle \pm |11\rangle) \nonumber \\
    \label{eq:bell}
    \frac{1}{\sqrt{2}}(|01\rangle \pm |10\rangle)
\end{align}

We leave it to those interested to convince themselves that one cannot write these states as a product of single qubit states.

\subsection{Choice of Basis}

As exposed in the Subsection \ref{ssec:info}, the operations that are allowed onto qubits can be expressed in terms of Pauli matrices. First, let us remind ourselves that, with the exception of $I_2$, which commutes with all the other Pauli matrices, none of the latter commute with each other:

\begin{equation*}
    [\sigma_x, \sigma_y] = 2i\sigma_z, \quad [\sigma_y, \sigma_z] = 2i\sigma_x, \quad [\sigma_z, \sigma_x] = 2i\sigma_y
\end{equation*}
which has important consequences in how measurement is performed in quantum computing (remember that we cannot measure all spin operators simultaneously!).

The following are the normalized eigenvectors of the Pauli matrices, where the sign comes from the fact that the corresponding eigenvalues are $\pm 1$:

\begin{align}
    \label{eq:eigenvectors}
    |z_{+}\rangle = \begin{pmatrix} 1 \\ 0 \end{pmatrix}, &\quad |z_{-}\rangle = \begin{pmatrix} 0 \\ 1 \end{pmatrix} \nonumber \\
    |x_{+}\rangle = \frac{1}{\sqrt{2}}\begin{pmatrix} 1 \\ 1 \end{pmatrix}, &\quad |x_{-}\rangle = \frac{1}{\sqrt{2}}\begin{pmatrix} 1 \\ -1 \end{pmatrix} \\
    |y_{+}\rangle = \frac{1}{\sqrt{2}}\begin{pmatrix} 1 \\ i \end{pmatrix}, &\quad |y_{-}\rangle = \frac{1}{\sqrt{2}}\begin{pmatrix} 1 \\ -i \end{pmatrix} \nonumber 
\end{align}

Looking at Equation \ref{eq:eigenvectors}, we notice that the canonical basis for the space of states of a single qubit is the eigenbasis of $\sigma_z$, and is often referred to as the computational basis, allowing even the eigenvalues of $\sigma_x$ and  $\sigma_y$ to be described as simple linear combinations of $|0\rangle$ and  $|1\rangle$. Another relevant identity we should call attention to involves the eigenvectors of $\sigma_x$, which we will elaborate further later on:

\begin{align}
    |+ \rangle = \frac{1}{\sqrt{2}}\begin{pmatrix} 1 \\ 1 \end{pmatrix} &= \frac{1}{\sqrt{2}}(|0\rangle + |1\rangle) \\ 
    |- \rangle = \frac{1}{\sqrt{2}}\begin{pmatrix} 1 \\ -1 \end{pmatrix} &= \frac{1}{\sqrt{2}}(|0\rangle - |1\rangle) 
\end{align}

\subsection{Multi-qubit states}

So far in this section we have dealt exclusively with the salient features of a single qubit. However, a single qubit is hardly interesting and certainly cannot provide any of the expected performance gains over classical computers when it comes to quantum chemistry. Fortunately, qubits can be operated in conjunction. A $n$-qubit state can be defined as the tensor product of single-qubit states, e.g., the state where all qubits are found in the $|0\rangle$ state:

\begin{equation}
    |00\cdots 0\rangle = |0\rangle^{\otimes n}  = |0\rangle_1 \otimes |0\rangle_2 \otimes \cdots \otimes |0\rangle_n
\end{equation}

Likewise, we can apply an operator that only affects one or a subset of qubits. Let $U_k$ be a unitary operator that operates on the $k$-th qubit:

\begin{align*}
    U_k|00\cdots  0\rangle &= I_1I_2 \cdots U_k \cdots I_n|00\cdots  0\rangle \\
    &= |0\rangle_1 \otimes |0\rangle_2 \otimes \cdots U_k|0\rangle_k \cdots \otimes |0\rangle_n
\end{align*}
that is, only the $k$-th qubit is affected and to the other qubits it is as though the identity operation is applied.

\subsection{Quantum computing logic}
\label{ssec:logic}

The term quantum computing is very broad and may refer to several distinct operational paradigms. Before moving forward, it is important and necessary to clarify that this review will be limited to considering the so-called logic or gate-based quantum computing. This computing model retains many characteristics from standard classical computing, that is, most of the gates are direct analogues or try to reproduce operations that are routine for classical processors. There is also the notion of universal gate sets that would be possible to implement across all architectures that are grounded on this model, with the underlying premise that any circuit can be decomposed into a small predefined set of simpler gates. This paradigm has pros and cons but is the most widely adopted and certainly the most frequently used in quantum chemistry applications.

Let us start by defining a gate as a basic operation onto a set of qubits. The Pauli matrices form the generators for all possible quantum gates, and they can themselves, in turn, be seen as gates. Therefore, the Pauli matrices $\sigma_x$, $\sigma_y$, $\sigma_z$ are represented by the $X$, $Y$, $Z$ gates. The $Z$ gate is also known as phase-flip, as it changes the sign when applied to the $|1\rangle$ ($|0\rangle$ is unaffected by it). An important characteristic of the $X$ gate is that it maps the projection along $|0\rangle$ onto $|1\rangle$ and vice-versa, and is thus called bit-flip operator:

\begin{equation*}
    X|0\rangle  = |1\rangle, \quad X|1\rangle  = |0\rangle, \quad X\begin{pmatrix} \alpha \\ \beta \end{pmatrix} = \begin{pmatrix} \beta \\ \alpha \end{pmatrix}
\end{equation*}

One can easily check that these gates applied on the computational basis return either of these two vectors (apart from a phase factor). As we have pointed out before, one of the necessary ingredients for quantum speedup is superposition of states. This is most commonly accomplished via the introduction of one of the most important single-qubit gates, the Hadamard gate $H$:

\begin{equation}
    H = \frac{1}{\sqrt{2}}\begin{pmatrix} 1 & 1 \\ 1 & -1 \end{pmatrix}
\end{equation}

We can highlight three interesting properties of the Hadamard gate. The first is its ability to create a superposition of states $|0\rangle$ and  $|1\rangle$

\begin{equation}
    H|0\rangle = \frac{1}{\sqrt{2}}(|0\rangle+|1\rangle), \quad H|1\rangle = \frac{1}{\sqrt{2}}(|0\rangle-|1\rangle),
\end{equation}
the second is that it can change the computational basis into the ``plus or minus'' basis, also known as ``X''-basis, and vice-versa (see Equation \ref{eq:eigenvectors}). The third is that the operator that carries out the Hadamard gate is self-adjoint, which means that applying it once to a state represented in the computational basis changes it to the ``X''-basis and applying it once more on to this state brings it back to the computational basis. This property is crucial in carrying out measurements when the qubits in the final state are not found in the computational basis.

A typical task in solving chemical problems is to recast the molecular Hamiltonian as a spin Hamiltonian, where all terms are given as products of Pauli matrices, a more appropriate representation for most quantum hardware. Mapping the chemical Hamiltonian into its spin analog is usually done by identifying orbitals or states with qubits. This means we can associate single-qubit gates as operators that act on single orbitals or states, hence they cannot alone account for interaction between orbitals/states, which is responsible for electron correlation in the context of electronic structure. Thus, in order to go beyond a single particle/state picture, we need to institute two-qubit gates, at the least. Among those, the most frequent is the CNOT gate, also known as CX or controlled-NOT:

\begin{equation}
    \text{CNOT} = \begin{pmatrix} 1 & 0 & 0 & 0  \\ 0 & 1 & 0 & 0 \\ 0 & 0 & 0 & 1 \\ 0 & 0 & 1 & 0 \end{pmatrix}
\end{equation}

The CNOT is the quintessential example of an entangling gate. Another important feature of the CNOT is that it is a controlled gate, that is, its action on a target qubit depends on the state of another qubit, referred to as control. The CNOT works by applying an $X$ gate on to the target qubit if the control qubit is found in the $|1\rangle$ state. Analogously, the CZ is also a relatively common controlled two-qubit gate where the target qubit undergoes a $Z$ gate operation. In order to understand precisely how this gate works, one needs to establish how to conveniently order the basis states. Starting with the Hadamard gate as an example, it should be clear that we can associated the row and columns with the basis states as such:

\begin{equation}
    H = \frac{1}{\sqrt{2}}\begin{pmatrix} 1 & 1 \\ 1 & -1 \end{pmatrix}
        \begin{matrix}
        |0\rangle \\ |1\rangle
    \end{matrix}
\end{equation}

However, the reader may be now wondering what to label the basis states if we have many qubits. Conventionally, the basis states are numbered 0 through $2^N-1$, where the bit string of the corresponding label is the $N$-bit binary representation of that number. Taking the CNOT gate to illustrate this, we have for states labeled by 0, 1, 2, and 3, which can be readily associated $|00\rangle$, $|01\rangle$, $|10\rangle$, and $|11\rangle$, in this order

\begin{equation}
    \text{CNOT} = \begin{pmatrix} 1 & 0 & 0 & 0  \\ 0 & 1 & 0 & 0 \\ 0 & 0 & 0 & 1 \\ 0 & 0 & 1 & 0 \end{pmatrix}
        \begin{matrix}
        |00\rangle \\ |01\rangle \\ |10\rangle \\ |11\rangle
    \end{matrix}
\end{equation}

Note that it is also possible to traverse the series by going from the left-most to the right most bit. The former option is called least significant bit (LSB), and is adopted most often, while the latter is referred to most significant bit (MSB).

The gates discussed above are not meant to be an exhaustive list of quantum logic gate operations, but to briefly introduce the ones most commonly found in quantum chemistry applications. There are gates that act on more than two qubits, e.g., Toffoli gate. Also, some gates despite not being necessary in describing a given algorithm, become important in their concrete implementation due to hardware demands. An example of this is the SWAP gate, a two-qubit gate the swaps two qubits, which can be necessary in certain devices that can only perform certain operations when the qubits onto which it operates on are adjacent to one another due to qubit connectivity limitations.

Table \ref{tab:gates} summarizes the most important gates discussed here.

\begin{table}[]
    \centering
    \caption{Most common gates in quantum algorithms for quantum chemistry and different ways they can be represented.}
    \begin{tabular}{c c c c}
    \hline
         Gate & Matrix representation & Operator representation (computational basis) & Circuit representation  \\ 
         \hline \\ 
         $X$ & $\begin{pmatrix} 0 & 1 \\ 1 & 0 \end{pmatrix}$ & $|1\rangle\langle 0| + |0\rangle\langle 1|$ & \begin{tikzcd} \qw & \gate{X}  & \qw \end{tikzcd} \\ [1cm]
         $Y$ & $\begin{pmatrix} 0 & -i \\ i & 0 \end{pmatrix}$ & $i|1\rangle\langle 0| - i|0\rangle\langle 1|$ & \begin{tikzcd} \qw & \gate{Y} & \qw \end{tikzcd} \\ [1cm]
         $Z$ & $\begin{pmatrix} 1 & 0 \\ 0 & -1 \end{pmatrix}$ & $|0\rangle\langle 0| - |1\rangle\langle 1|$ & \begin{tikzcd} \qw & \gate{Z} & \qw \end{tikzcd} \\[1cm]
         $R_x(\theta)$ & $\begin{pmatrix} \text{cos}(\frac{\theta}{2}) & -i\text{sin}(\frac{\theta}{2}) \\ -i\text{sin}(\frac{\theta}{2}) & \text{cos}(\frac{\theta}{2}) \end{pmatrix}$ & $\text{cos}(\frac{\theta}{2})(|0\rangle\langle 0| + |1\rangle\langle 1|) - i\text{sin}(\frac{\theta}{2})(|1\rangle\langle 0| + |0\rangle\langle 1|)$ & \begin{tikzcd} \qw & \gate{R_x(\theta)} & \qw \end{tikzcd} \\ [1cm]
         $R_y(\theta)$ & $\begin{pmatrix} \text{cos}(\frac{\theta}{2}) & -\text{sin}(\frac{\theta}{2}) \\ \text{sin}(\frac{\theta}{2}) & \text{cos}(\frac{\theta}{2}) \end{pmatrix}$ & $\text{cos}(\frac{\theta}{2})(|0\rangle\langle 0| + |1\rangle\langle 1|) +  \text{sin}(\frac{\theta}{2})(|1\rangle\langle 0| - |0\rangle\langle 1|)$ & \begin{tikzcd} \qw & \gate{R_z(\theta)} & \qw \end{tikzcd} \\  [1cm]
         $R_z(\theta)$ & $\begin{pmatrix}e^{-i\frac{\theta}{2}} & 0 \\ 0 & e^{i\frac{\theta}{2}} \end{pmatrix}$ & $e^{-i\frac{\theta}{2}}|0\rangle\langle 0| + e^{i\frac{\theta}{2}}|1\rangle\langle 1|$ & \begin{tikzcd} \qw & \gate{R_z(\theta)} & \qw \end{tikzcd} \\ [1cm]
         $H$ & $\frac{1}{\sqrt{2}}\begin{pmatrix} 1 & 1 \\ 1 & -1 \end{pmatrix}$ & $\frac{1}{\sqrt{2}}(|0\rangle\langle 0| + |0\rangle\langle 1| + |1\rangle\langle 0| - |1\rangle\langle 1|)$ & \begin{tikzcd} \qw & \gate{H} & \qw \end{tikzcd} \\ [1cm]
         CNOT & $\begin{pmatrix} 1 & 0 & 0 & 0  \\ 0 & 1 & 0 & 0 \\ 0 & 0 & 0 & 1 \\ 0 & 0 & 1 & 0 \end{pmatrix}$ & $I_1\otimes I_2( |00\rangle\langle 00| + |01\rangle\langle 01|) + I_1 \otimes X_2 ( |11\rangle\langle 10| + |10\rangle\langle 11|)$ & \begin{tikzcd} \qw & \ctrl{1} & \qw \\ \qw & \targ{} & \qw \end{tikzcd} \\ [1cm]
         \hline
    \end{tabular}
    \label{tab:gates}
\end{table}

\subsection{Mapping fermions to qubits}
\label{ssec:mapping}

Foundational to the electronic structure theory is the fact that electrons are fermions. That carries profound consequences, one of which is the antisymmetry between fermionic degrees of freedom. Practically, this means that permutation of the coordinates of any two fermions leads to a change in sign of the underlying wave function. We now remind the reader that a convenient formulation of the electronic structure problem is the second quantization,\cite{jorgensen, shavitt_bartlett_2009} where the wave function is represented by a state in the full Fock space, which considers all possible particle numbers. This is also referred to as the occupation basis, where the full wave function can be represented by a tensor product of one-particle wave functions that can be either occupied ($|1\rangle$) or unoccupied ($|0\rangle$). And it is this binary outcome regarding the occupation of the one-particle wave functions, i.e., orbitals, that facilitates the mapping of fermionic modes into spin degrees of freedom for quantum hardware simulation.

Once the electronic structure problem is represented in terms of second quantized operators, a proper transformation into the qubit counterparts is called for since native spin operators (Pauli matrices) do not account for the fermionic antisymmetry. The most widely used such approaches are Jordan-Wigner (JW),\cite{JW} Bravyi-Kitaev (BK),\cite{bk1, bk2, bk3, bk4} Parity, and Bravyi-Kitaev SuperFast (BKSF),\cite{bksf} while more adaptable strategies also exist.\cite{Steudtner_2018} These mapping techniques are isomorphic with the original fermionic formulation, which means that they do not affect the expectation value of the desired observables. Yet, this fermion-to-qubit mapping is not unique, and even though it may be nothing more than an afterthought in devising algorithms from a purely chemical standpoint, choosing an appropriate mapping can have drastic performance consequences when deploying simulations to the currently available devices. As a brief example, the JW transformation is the most common, due to be known for a long time and also because its simplicity. Yet, the way antisymmetry is accounted for leads to a high degree of non-locality, by which we mean that the operator associated with a single spin-orbital needs to act jointly on many qubits. In general, the operator for the $p$-th orbital will act on all qubits 1, 2, $\dots$, $p$, and particular to the JW instance, it implies in the inclusion of $Z$ gates for all $p-1$ preceding qubits. This makes the JW mapping scale as $\mathcal{O}(N)$ in the number of qubits, which is the same scaling observed for the Parity mapping. The BK counterpart, on the other hand, is not as straighforward, yet much more compact and local, scaling as $\mathcal{O}(\text{log}_2N)$.

\subsection{Quantum measurement}
\label{ssec:measurememt}

The laws of quantum mechanics state that we cannot directly observe a state/wave function, but are restricted to drawing conclusions upon measurements of observables, and that this measurement in turn collapses the state onto one of the eigenstates of the measured observable. Closely related in quantum information is the famous \emph{no-cloning theorem}, which states that it is impossible to make copies of quantum states.\cite{Wootters1982, Dieks1982} This is an immediate consequence of how quantum measurement works, and it has drastic consequences for quantum computing. In simple terms, the measurement of an observable with operator representation $O$ in a discrete space is given by the expectation value of $O$:

\begin{equation}
    \label{eq:measurement}
    \langle O \rangle = \frac{\langle \psi|O|\psi\rangle}{\langle \psi |\psi \rangle}
\end{equation}

Unless there is certainty that $|\psi \rangle$ has been prepared as an eigenstate of $O$, a series of measurements take place in order to probe the distribution of eigenstates that compose $|\psi \rangle$, with each of the eigenvectors being associated a probability of being measured. We can represent the state by  $|\psi \rangle = \sum_k o_k|o_k\rangle$, with probability $p_k$ that the state will collapse onto $|o_k\rangle$. Projecting $|\psi \rangle$ along $|o_k\rangle$ with the projector $P_k = |o_k\rangle \langle o_k|$ leads to:

\begin{equation}
   \langle P_k \rangle =  \frac{\langle \psi |o_k\rangle \langle o_k|\psi \rangle}{\langle \psi |\psi \rangle} = \frac{|\langle o_k|\psi\rangle|^2}{\langle \psi |\psi \rangle}
\end{equation}

Stating that $\sum_k|o_k\rangle \langle o_k| = I$, and assuming a normalized states ($\{|o_k\rangle\}$ are orthogonal by virtue of $O$ being Hermitian), it follows that:

\begin{equation}
    \langle \sum_k P_k \rangle = \langle \psi|\left(\sum_k |o_k\rangle \langle o_k| \right)|\psi\rangle = \sum_k|\langle o_k|\psi\rangle|^2  = 1 = \sum_k p_k 
\end{equation}
where in the last equality we evoke $\sum_k|o_k\rangle \langle o_k| = I$.

Thus one can identify the probabilities $p_k$ as the probability of the state collapsing onto $|o_k\rangle$ with the norm squared of the projection along corresponding eigenvector. It also follows that Equation \ref{eq:measurement} can be readily expressed in terms of probabilities:

\begin{equation}
    \langle O \rangle = \sum_k o_kp_k
\end{equation}

It is important to note here that, although these are basic principles that guide quantum measurement in general, qubit states do not comprise a perfect closed system, thus projective measurements like the ones shown here are largely pedagogical and, even though they serve as the basis for how measurement is performed in quantum computers, operation of these devices require more elaborate strategies -- the projectors $P_k$ can be replaced by more general measurement operators which still follow the basic premises above, but are not necessarily unitary. Also, one needs to make a choice regarding the basis in which they measurement will take place, with the $Z$ basis being an obvious choice. In order to be able to measure the products of Pauli operators that comprise the Hamiltonian, the circuit may be appended by gates that perform rotations to the computational basis (remember that by convention the measurements are taken as projections onto the $z$-axis of each qubit). That being said, a $Z$ operator is \emph{diagonal} when measured in the computational basis, meaning it does not require any further gates. The $X$ and $Y$, on the other hand, need to be brought to the $Z$ basis. Only after performing this change of basis the measurement itself takes place.

\subsection{Quantum circuits}
\label{ssec:circuits}

As it is common knowledge in quantum chemistry, one could hardly recover the ground state of a many-electron system by a single many-body rotation of an uncorrelated, mean-field starting point, while it would also be difficult to expect that a single time step would enable one to unravel the entire dynamics of a quantum system. Analogously, and by now the connection between quantum mechanical operators and their effect on to quantum states should be more apparent, if starting from a mean-field approximation and having one- and two-qubit gates at one's disposal, one should expect that a series of such gates will be needed in order to achieve a sound description of many-body states of the type relevant in quantum chemistry.

\begin{figure}[]
\resizebox{\columnwidth}{!}{
    \centering
  \begin{quantikz}
  \lstick{\textbf{a)}} & & & & \lstick{\textbf{b)}} & & & & & \lstick{\textbf{c)}} & & \\
&\lstick{$\ket{0}_0$}& \qw & & &\lstick{$\ket{0}_0$} & \gate{X} & \qw & & &  \lstick{$\ket{0}_0$} & \gate{X} & \gate{R_x(\pi/2)} & \ctrl{1} & \qw & \qw & \qw & \qw & \qw & \ctrl{1} & \gate{R_x(-\pi/2)} & \qw\\
& \lstick{$\ket{0}_1$}& \qw & & &\lstick{$\ket{0}_1$} & \qw  & \qw  & & & \lstick{$\ket{0}_1$} & \qw & \gate{H} & \targ{} & \ctrl{1} & \qw & \qw & \qw & \ctrl{1} & \targ{} & \gate{H} & \qw \\
& \lstick{$\ket{0}_2$}& \qw & & &\lstick{$\ket{0}_2$} & \gate{X} & \qw  & & & \lstick{$\ket{0}_2$} & \gate{X} & \gate{H} & \qw & \targ{} & \ctrl{1} & \qw  & \ctrl{1} & \targ{} & \qw & \gate{H} & \qw\\
& \lstick{$\ket{0}_3$}& \qw & & &\lstick{$\ket{0}_3$} & \qw & \qw      & & & \lstick{$\ket{0}_3$} & \qw & \gate{H} & \qw & \qw & \targ{} & \gate{R_z(\theta)} & \targ{} & \qw & \qw & \gate{H} & \qw
 \end{quantikz}
 }
    \caption{Sample circuit representation of the a) physical vacuum ($|\Psi\rangle = |0000\rangle$); b) Hartree-Fock state for H$_2$ in a minimal basis, where the bits corresponding to the two lowest-energy spin-orbitals are flipped via the $X$ gate (($|\Psi\rangle = X_0X_2|0000\rangle = |1010\rangle$ )); c) the superposition of the HF state and the determinant from the double excitation upon the action of $ e^{-i\theta Y_0X_1X_2X_3/2}$ ($|\Psi\rangle = e^{-i\theta Y_0X_1X_2X_3/2}|1010\rangle = c_0|1010\rangle + c_1|0101\rangle$)}
    \label{fig:circuit}
\end{figure}
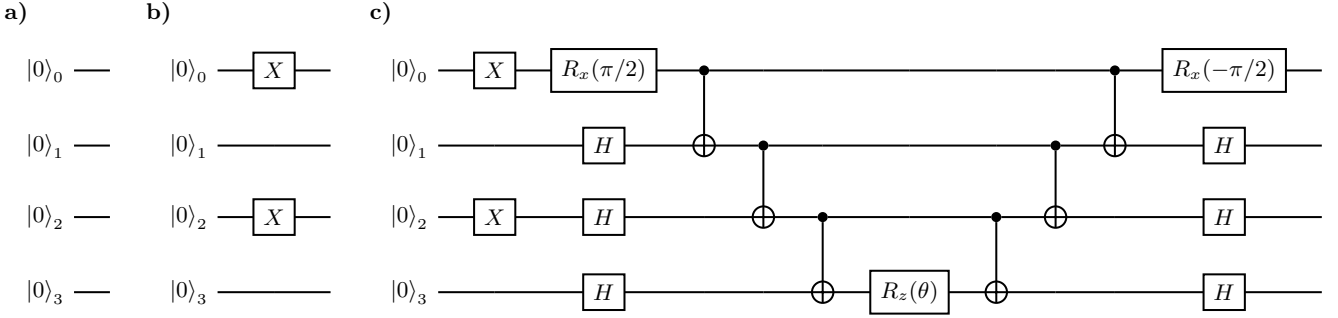

As discussed in Subsection \ref{ssec:logic}, the type of quantum computation we are considering here is often referred to as logic or gate-based due to the effort of retaining a good deal of what has long been established in classical computation. Since we are discussing the operations of a computer, it seems cumbersome and undesirable to deal with these operations purely from the perspective of the algebra of the respective quantum mechanical operators. Of course, the underlying quantum mechanics is always there, but turning to the routine circuit notation makes the connection to computer logic transparent and is often more abbreviated and readily understandable. Moreover, it also carries the benefit of making explicit certain aspects that are relevant in terms of concrete implementation, such as the gates used and the circuit depth, and thus should be of little surprise that this diagrammatic notation has long been embraced by the quantum computing community. In order to illustrate the importance of these diagrams and to help develop familiarity with these constructs, we walk the reader through some of the most relevant aspects of quantum circuits, which is displayed in Figure \ref{fig:circuit} and whose building blocks are found in Table \ref{tab:gates}.

Let us take the H$_2$ in the STO-3G basis set\cite{sto3g1} as an example to illustrate the construction of a quantum circuit and how it prepares the corresponding quantum state. We start by assuming the most straightforward encoding of the electronic Hamiltonian, as provided by the Jordan-Wigner transformation (See Sec. \ref{ssec:mapping}). In second quantization, the operators associated to spin-orbitals are mapped onto Pauli matrices for the respective qubits in a one-to-one fashion. Thus, for a $N$-spin-orbital Hamiltonian, whose corresponding creation operators are $\{a_1^\dagger, \dots , a_N^\dagger\}$ and annihilation operators $a_p = (a_p^\dagger)^\dagger$, it would required $N$ qubits. In passing, we mention that more sophisticated strategies can also be employed here with the advantage of yielding orbital operators that are more local (involve fewer qubits), potentially furnishing smaller, more-local operator product in the qubit basis, and potentially even lowering the underlying qubit count. Because in the most typical case we are dealing with a non-relativistic, spin-free Hamiltonian, each spin-orbital $\psi$ is separable in terms of its spatial and spin components. Mathematically, this means $\psi(\mathbf{r}; s) = \phi(\mathbf{r})\omega(s)$, where $\phi(\mathbf{r})$ is given by one of the many atomic basis sets available,\cite{bse} and allowing construction of two spin-orbitals ($\omega(s)$ = $\alpha$ and $\beta$ or $\uparrow$ and $\downarrow$ spins) out of the same spatial function. One mapping convention indexes two adjacent qubits to a pair of $\alpha$ and $\beta$ spin-orbitals from the same spatial orbitals, while another indexes all $\alpha$ spin-orbitals in the first $N/2$ qubits then all $\beta$ in the remaining ones, with the same spatial orbital ordering within the group of qubits with the same spin.

Assuming the widely adopted convention that the qubit states $|0\rangle$ and $|1\rangle$ represent unoccupied and occupied orbitals, respectively, the qubit register is initialized at $|0\dots 0\rangle$, representing the physical vacuum, and illustrated in Figure \ref{fig:circuit}a. The horizontal lines next to each qubit are called ``wires'' and show the gates which are applied to its corresponding qubit. Common practice is to proceed by preparing a Hartree--Fock (HF), which can be easily carried out in the JW representation by flipping the bits, through the $X$ gate, in the qubits associated with the $N_e$ lowest energy spin-orbitals ($N_e$ being the number of electrons). In our case, these would be qubits 0 and 2, as shown in Figure \ref{fig:circuit}b.


The circuit represented by Figure \ref{fig:circuit}b is capable of preparing a HF state, which can be computed efficiently by classical computers; hence a more complex circuit is needed if the quantum computer is to provide any advantage. The expectation is that quantum computers will be efficient in preparing entangled many-body states, which in the present context means accounting for electron correlation. As said previously, this cannot be achieved without operations that simultaneously act onto more than a single orbital/state. In typical electronic structure, this means that we need at least double excitations out of the reference function. In second quantization, these are given by operators of the form $a^\dagger_a a_i a^\dagger_b a_j$, meaning electrons are annihilated in the occupied orbitals $i$ and $j$ and created in the virtual orbitals $a$ and $b$. Analogously, such an operation, when properly encoded in terms of spin operators, will involve at least four qubits, possibly more due to the introduction of trailing $Z$ operators that ensure proper fermionic antisymmetry when we have more than four qubits. Due to how the transformations in Sec. \ref{ssec:mapping} map the second quantized operators onto Pauli operators, a double excitation operator can be identified with a product of operators of the general form $(X_i \pm iY_i)(X_j \pm iY_j)(X_k \pm iY_k)(X_l \pm iY_l)$, where there would be a number of prepended $Z$'s in the most general case. Such a product of Pauli operators are referred to in the literature also as Pauli words or strings. Thus, we are left with a sum of Pauli strings that can be as simple as a product of four operators for a double excitation operator. Keeping in mind that a gate that implements the Pauli word $P$ can be written in general form as $\text{exp}(-i\theta P/2)$, for $\theta \in \mathbb{R}$, we need an odd number of Pauli $Y$'s. This requirement stems from the fact that each $Y$ operator contributes a imaginary unit $i$, along with the $i$ in the definition of the quantum gate, thus in order to be able to measure an observable, whose expectation value is by definition a real number, the total number of $i$'s must be even. This implies that, the surviving terms have the form of either $Y_iX_jX_kX_l$ or $Y_iY_jY_kX_l$ and the corresponding permutations. Taking $Y_0X_1X_2X_3$ as an example and applying it to our HF state, we are left with Figure \ref{fig:circuit}c.

We can see that some of the gates we mentioned before are easily identified in the circuits of Figure \ref{fig:circuit}, e.g., $X$ and $H$ gates. Gates such as $R_z(\theta)$ mean a rotation by $\theta$ around some axis, in this case $z$, and can expressed as $R_z(\theta) = \text{exp}(-i\theta Z / 2)$. Analogously, the $R_x(\pi/2) = \text{exp}(-i\pi Z / 4)$ and $R_x(-\pi/2) = \text{exp}(i\pi Z / 4)$ gates acting on qubit 0 can be pictured as a rotation of $\pi/2$ and $-\pi/2$ around the $x$-axis in the Bloch sphere for that qubit. It should be clear from our discussion in Subsection \ref{ssec:logic} and by visual inspection of Figure \ref{fig:circuit} that these gates are single-qubit gates, evidenced by being restricted to a single wire. The gates that span two wires in Figure \ref{fig:circuit} are the circuit representation of the CNOT gate. The $\bullet$ is placed on the wire corresponding to the control qubit and the $\oplus$ is therefore placed on the target qubit, with the vertical line connecting the two qubits involved. Only the qubits where the control and target symbols are found are subject to this gate (or any two-qubit gates, with different symbols), and it is important to point out because the CNOT is not necessarily restricted to adjacent qubits. This pattern of CNOTs covering adjacent qubits is colloquially named CNOT ``ladder'' and often emerges in circuits for quantum chemistry simulations, being a well-known signature of the JW encoding.

From a quantum hardware perspective, the implementation of quantum circuits is very much alike a spectroscopy experiment, with the varying specifics depending on the technology behind the chip architecture, e.g., superconducting, ion-traps, etc. Each gate requires a length of time to be implemented, and although knowing exactly how long it takes for the quantum computer to implement each gate is not crucial at this point, it is important to understand that this implementation takes place in a time-ordered series, from left to right in the circuit. Extending the circuit with more gates along a given wire makes the circuit \emph{deeper}, with the wire containing the most gates being the one which defines the depth of the circuit. This is a crucial metric because it relates to the coherence time of the qubits in the device. The deeper the circuit, the more likely its implementation will suffer from the deleterious effects of decoherence, which prompts the devising of shallow circuits for successful execution in NISQ hardware.

Last but not least, the circuit in Figure \ref{fig:circuit}c is only responsible for preparing the state $|\Psi\rangle = c_0|1010\rangle + c_1|0101\rangle$, thus lacking the measurement instructions (Section \ref{ssec:measurememt}). If we think of a typical electronic Hamiltonian, it is comprised of $\mathcal{O}(N^4)$ individual operators for a basis set of $N$ orbitals, each of which needs to be measured separately. Since we often do not intend to prepare eigenstates of these terms, but rather of the entire Hamiltonian, measurements of these terms will produce a distribution over the possible eigenstates, hence requiring many measurements of each individual term in order to achieve a satisfactory picture of this distribution. In this process, the same circuit needs to be implemented \emph{from scratch} every time, a consequence of quantum measurement. In each of these repetitions, also known as ``shots'', the state collapses onto one of eigenstates of the Hamiltonian term being measured. Thus, if $N_\text{shots}$ shots are expected to probe the distribution of states to a desired precision when measuring an observable with $N_\text{terms}$, then the circuit is implemented $N_\text{terms} \times N_\text{shots}$ times in order to estimate the observable in question.

\subsection{Trotter-Suzuki expansion and time evolution}

Recapitulating our discussion in Subsection \ref{ssec:circuits}, much of the features sought when turning to quantum computers, such as entanglement, come from two-qubit gates (or multi-qubit gates more generally). On the other hand, their implementation is accompanied with many technical challenges, hence the attempt to avoid them. The example illustrated by Figure \ref{fig:circuit} and the corresponding problem does not show that Hamiltonians in bases with more orbitals result in the spin operators that can span many qubits. In other words, it can generate a very \emph{non-local} representation of the operator in question, which can be partially mitigated depending on the choice of mapping. In the case of the unitary coupled cluster\cite{ucc1, ucc2, ucc3} with single and double excitations (UCCSD), this would mean having to implement the entangler from the exponentiation of $\sum_{ia}a^\dagger_a a_i + \sum_{ijab}a^\dagger_a a_ia^\dagger_b a_j - \text{h.c}$, which would lead to a gate spanning all qubits in the register. In order to circumvent this difficulty, one can recognize that this unitary can be recast with the aid of the Trotter-Suzuki approximation:\cite{Hatano2005}

\begin{equation}
\label{eq:trotter}
    e^{A+B} = \lim_{n\rightarrow \infty} (e^{A/n}e^{B/n})^{n}
\end{equation}
for $A$ and $B$ non-commuting operators.

The equality in equation \ref{eq:trotter} holds only for $n\rightarrow \infty$, which is obviously unfeasible. The alternative is to resort to low-orders in the Trotter-Suzuki expansion (small $n$), which for the sort of operator we are to deal with in quantum chemistry is the following to first-order

\begin{equation}
    e^{\sum_k \theta_k (T_k-T^\dagger_k)} = \prod_k e^{\theta_k (T_k-T^\dagger_k)} + \mathcal{O}(\sum_k \theta^2_k)
\end{equation}

Many applications of the UCCSD in the literature, or many other ansatze that depend on a collection of non-commuting operators,  use the Trotter-Suzuki formula, or the ``trotterized'' ansatz, in low-orders. 

The Trotter-Suzuki formula is recurring in the context of quantum algorithms for time evolution, arguably one of the most anticipated use cases for quantum computing in the physical sciences. Ideally, a quantum computer can naturally lend itself to such a task by recognizing the time propagator $e^{-iHt}$ as a unitary operator and its translation into logic operations (gates). The total propagation time can be parceled out into $n$ time steps, as in Equation \ref{eq:trotter}. Although conceptually straightforward, it is often necessary to take a large number of small time steps to achieve an accurate description of the time evolution. Such a description is synonymous with acceptable precision, which comes at the expense of extremely deep circuits, well beyond NISQ capabilities. The exponential difficulty in evolving states through time is compounded by the fact that the Hamiltonians involved may be highly non-local and non-sparse. Thus, routine exploration of the dynamics of chemical systems is a topic expected to be only viable with the advent of fault-tolerant quantum computers.

\section{Current landscape and state-of-the-art}

As hinted throughout this review, what is at the moment perceived as the current state-of-the-art in quantum computation is severely curtailed by the intrinsic limitations found in NISQ devices. Such processors have a relatively small number of qubits, with the current record being held by IBM with its Eagle 127-qubit processor.\cite{eagle} Other noteworthy examples are Google's Sycamore (53 qubits)\cite{sycamore} and the previous IBM's front runners, the Hummingbird family (65 qubits).\cite{hummingbird} The coherence among these qubits does not endure for long periods of time, translating into a serious cap in the depth of the circuits that the current hardware can implement. It is important to emphasize that the number of qubits alone should not be taken as a metric for performance. Alternative and complementary ways to gauge performance and fit for specific purposes have been introduced, such as quantum volume,\cite{quantum_volume} and Circuit Layer Operations Per Second (CLOPS).\cite{clops} Those will not be further explored here.

Currently, the circuits that can be realistically considered are far too shallow for many chemical applications. These machines are notorious for the inherent noise in several processes in their operation, which have proven quite challenging to model and control. However, single-qubit gates are quite efficiently implemented, with error rates comparatively low, which are often orders of magnitude below the total error accrued along the entire circuit. For that reason, single-qubit gates are often seen as \emph{error-free}, at least in the scale of the other errors introduced that creep in during the operation of these devices. Other, multi-qubit gates are more problematic, but as previously mentioned, are absolutely crucial to allow more sophisticated methodologies. The conspicuous presence of noise implies that experiments must be replicated in order to circumvent or alleviate many problems, for instance, the presence of unwanted local minima in the context of multi-variate optimization, such as in ansatze with many variational parameters (See Section \ref{ssec:circuits}). Another problem in this regard comes from the fact that qubits are not always isolated two-state systems, but are often two states out of a whole spectrum, which are not perfectly isolated from the remaining ``unwanted'' states, potentially leading to `leakage'' out of the computational subspace. In passing, quantum computation can be performed with units of information where more than two states are superimposed, which are called qudits in general.\cite{qudit} Technologies based on such manifold of states are not as mature as qubits and only timid advances have been reported with potential chemistry applications.\cite{qudit_chem1, qudit_chem2}

Due to these limitations, it is desirable to restrict the usage of current quantum computers to only those tasks which they can be potentially superior to their classical counterparts. This constraint has led to the proposal of hybrid compute strategies, coupling heterogeneous architectures, in this case classical and quantum computers, where the functionality of the former is augmented by envisioning the latter as accelerators or co-processors. This has given rise to the so-called hybrid quantum-classical paradigm, and unless stated otherwise, quantum computing of today falls almost exclusively within the purview of this archetype. On one hand, hybrid quantum-classical computing can potentially explore certain computational tasks deemed intractable by classical computers alone. On the other hand, it opens the door to other challenges, such as concerted operation across different architectures, with widely different data structures, programming languages, and the nature of the information being manipulated. 

At this point, we are going to provide a \textit{very} brief exposition on some of the most promising instances of quantum computing with relevance to quantum chemistry. This is not meant to be an exhaustive list of applications nor should it be expected to go into detail on the specifics of each case considered here. Readers interested in such material are referred to the related reviews on Algorithms and Devices.\cite{devices}

With the idea of a quantum co-processor in mind, the classical computer is still in charge of a substantial part of the workflow as a whole, that is, reading in and processing classical data, e.g., the Hamiltonian matrix elements and the representation of the corresponding circuits for their measurements. The quantum processor serves as a back-end, and can assume a virtual or a physical, concrete form. While one can refer to a quantum chemistry experiment performed in quantum hardware as a quantum simulation, this term can carry some ambiguity as one can ``simulate'' the simulation. Virtual back-ends enable simulation of what would be expected should an experiment be deployed to an actual quantum computer. A virtual back-end is a classical program that mimics the operation of a quantum computer by building the mathematical manipulations found in the latter, i.e., vector manipulations in some 2$^N$ Hilbert space. As such, these programs are expected to do precisely what make classical computers fall short of what quantum simulations demand in the first place, so they are limited in scope. However, it can offer valuable insights in several aspects, such as providing resource requirements estimates and bounds to quantities being computed. 

\begin{figure}
    \centering
    \includegraphics[width=\columnwidth]{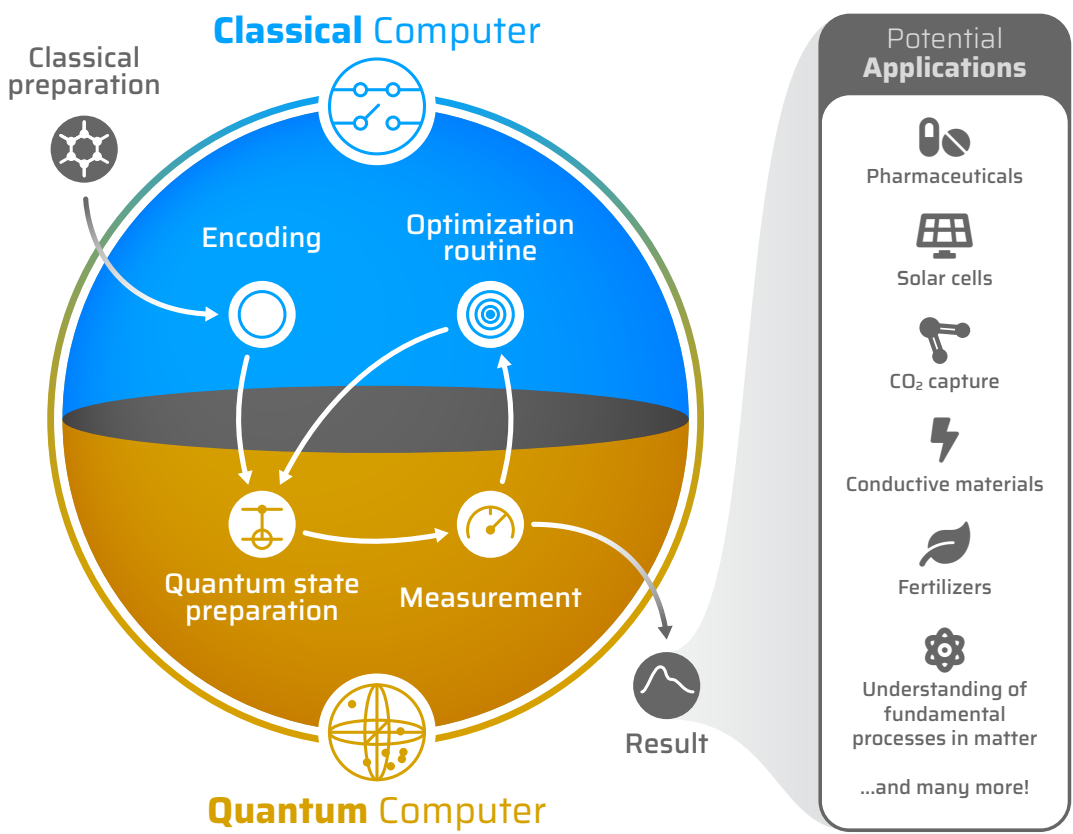}
    \caption{A pictorial representation of hybrid quantum-classical algorithms, VQE in particular, and some of its potential chemical applications.}
    \label{fig:vqe}
\end{figure}

Previously, it was mentioned that an eigenvalue can be recovered from a spectrum by the rather general quantum phase estimation algorithm. Notwithstanding its ability to achieve arbitrary precision, it would require fault-tolerant quantum computers for most applications due to the extremely deep circuits it produces and the number of qubits necessary for such a level of precision. A solution to this problem, which has emerged as the prime example of a hybrid algorithm and is the bedrock of many quantum algorithms in the NISQ era is the variational quantum eigensolver (VQE).\cite{vqe} Building on the variational principle, one assumes a trial wave function, which is implemented via a circuit with variable parameters, such as $\theta$ in the $R_z$ gate of Figure \ref{fig:circuit}c. These parameters are varied with the goal of minimizing the expectation value of the Hamiltonian, which is an upper bound to the ground state by virtue of the variational principle. For a given set of parameters, the quantum computer prepares the state and carries out the necessary measurements for each one of the individual terms in the Hamiltonian after it is transformed into a spin form (See Subsection \ref{ssec:mapping}). Once the energy is estimated for the current set of variational parameters, a classical optimizer checks for convergence, and updates the parameters if necessary, which in turn leads to a different state, and this cycle goes on until the optimizer meets some user-defined criteria of convergence. A graphic representation of the VQE algorithm showing the interplay of its basic components spanning the two disparate architectural paradigms is shown in Figure \ref{fig:vqe}.

A great deal of effort in the VQE algorithm for chemical simulation purposes is found in devising physically sound and computationally efficient ansatze, which comprises the state preparation stage. More often than not, these two desired features are at odds with one another. Different forms of the circuit ansatz may present themselves and here we point to some options that have been found in many reported applications. Many of these proposals are guided by the unitary coupled cluster\cite{ucc1, ucc2, ucc3} with single and double excitations (UCCSD).\cite{mcclean2016theory, romero2018strategies, particle_hole} Straightforward application of the UCCSD within VQE demands unfeasible multi-qubit gates, which is often circumvented by resorting to the Trotter-Suzuki approximation, while introducing the problem that the ansatz is dependent on the order the operators are included.\cite{trotter_uccsd} Alternative circuits that do not rely on Trotterization have also been proposed recently.\cite{jastrow, gate_fabric} 

\begin{figure}[!h]
    \centering
    \includegraphics[width=.75\columnwidth]{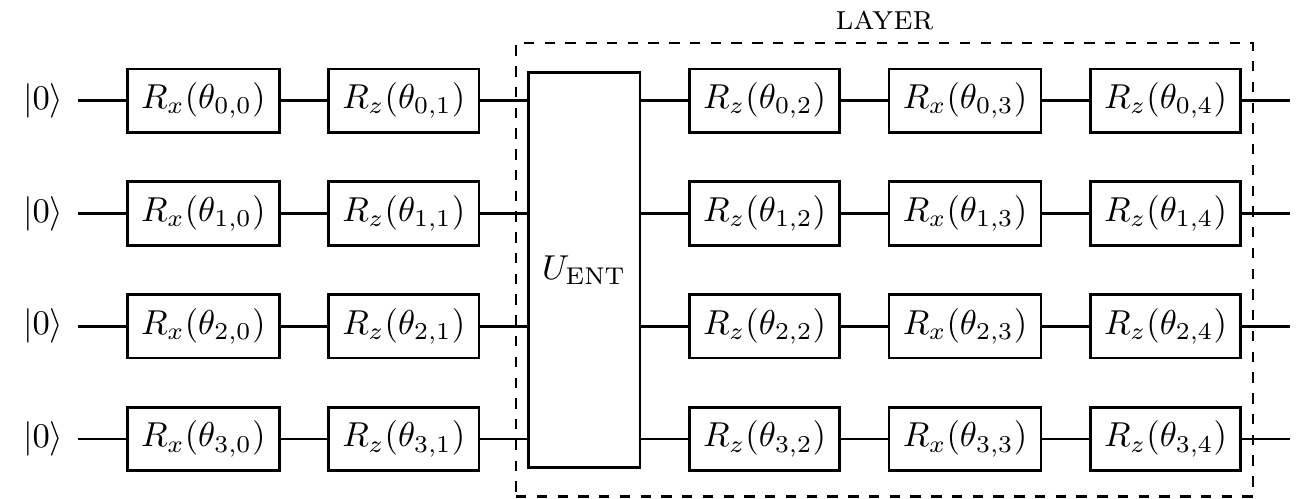}
    \caption{Depiction of the layered structure of a typical hardware-efficient ansatz. The ``LAYER'' unit can be repeated according to the application and the generic $U_\text{ENT}$ block is often comprised of two-qubit entangling gates, such as CNOT or cross-resonance gates.\cite{cr1, cr2}}
    \label{fig:hwe}
\end{figure}

The search for more amenable implementations of VQE and its extension to other domains has spurred many of its variations. Shallower circuits can in principle be obtained by the Adaptive Derivative-Assembled Pseudo-Trotter (ADAPT) ansatz coupled with the VQE protocol, but at the potential expense of many more measurements.\cite{adapt, qubit_adapt} In the special case of a collection of weakly-interacting chemical units, akin to certain photoactive complexes, the fermionic antisymmetry requirement can be relaxed and the ab initio exciton model (AIEM)\cite{aiem1, aiem2, aiem3} serves as an excellent approximation. By mapping electronic states instead of orbitals onto qubit states, the MC-VQE\cite{mc-vqe} (MC stands for multi-contracted) can achieve striking accuracy in the computation of absorption spectra, and this algorithm has also been extended to computation of forces and other response properties.\cite{mcvqe_gradients} Another noteworthy example is found when sacrificing the chemical interpretation provided by UCCSD for a more amenable circuit construction prescription, relying on the so-called hardware efficient ansatz,\cite{kandala2017hardware} which is depicted in Figure \ref{fig:hwe}. The efficiency gains come at the expense of heavily parameterized circuits whose accuracy can be tuned with the addition of ``layers'' of entanglers ($U_\text{ENT}$ in Figure \ref{fig:hwe}), which are groups of predefined gates generally containing CNOTs or other entangling gates. As more layers are added, the gain in accuracy is accompanied by adding to the depth of the circuit. The intrinsic difficult in finding the minima for such a circuit has been reported largely mitigated for some small molecules with the aid of global optimization based on genetic evolution approaches.\cite{mog-vqe} There has been a number of accounts proposing similar layered state preparation strategies with varying gates and rationales.\cite{kup, swap_networks, bassman2021constantdepth, gate_fabric}

Another family of methods that have emerged recently as a viable alternative to VQE-based algorithms can be grouped as stemming from time evolution considerations, with particular interest in imaginary time evolution (ITE).\cite{McArdle2019, motta2019determining} Quantum ITE (QITE) carries the prospect of delivering not only the eigenstates of the Hamiltonian, but also thermal states, which lend them naturally to the treatment of open systems. A similar avenue can be found in the quantum analog of the well-known power method,\cite{PRXQuantum.2.010333} which shares some of the theoretical foundations with ITE-type approaches. In a similar vein, quantum computers can be employed to prepare entangled states with sizable overlap with the true ground state and carry out the subsequent measurements in an efficient fashion. This can pave the way for widespread adoption of methodologies which estimate the ground state based on moments of the Hamiltonian,\cite{Kowalski2020, peng2021variational, Vallury2020quantumcomputed, claudino2021improving} which share some commonalities with ITE.

The development of quantum algorithms for chemistry has taken a similar path to classical approaches insofar as the effort has been primarily and largely devoted to the description of the ground state. This is not to say there has not been successful endeavors in terms of excited states, but that the search of suitable approximations for the ground state has dominated the picture. Here some of the proposed techniques to compute excited states are mentioned. VQE can still be used in this context, given the necessary modifications, which often include the introduction of penalty terms in the form of Lagrange multipliers to avoid entering the subspace where the ground state can be found.\cite{penalty, GreeneDiniz2020} As mentioned above, both MC-VQE and ITE can naturally treat excited states by construction, with the former being able to deliver absorption spectra in the limit where the AEIM assumptions are valid and the latter not being limited to the eigenspectrum of the Hamiltonian. Moreover, a particular case of the expansion of moments of the Hamiltonian, the Peeters-Devreese-Soldatov (PDS)\cite{peeters1984upper, soldatov1995generalized} energy functional can also yield upper bounds to excited states. Excited states can also be obtained by methods that are largely inspired by classical counterparts, namely the widespread equation of motion (EOM)\cite{eom1, eom2} and its quantum analog QEOM\cite{qeom} and the quantum subspace expansion,\cite{qse1, qse2} which borrows from the long-standing linear response\cite{lrcc} theories in quantum chemistry.

A common shortcoming shared by many of these strategies, and also found in many of the NISQ-prone algorithms, is the sheer number of measurements involved. Electronic Hamiltonians have $\mathcal{O}(N^4)$ terms for a basis with $N$ spin-orbitals, and the sampling in the present NISQ devices typically incurs $\mathcal{O}(10^3)$ independent measurements (shots) to achieve acceptable error rates. Thus, it is easy to see that even for small Hamiltonians, the associated overall number of times the quantum computers need to synthesize a circuit can be staggering. Many have been the angles from which this problem has been attacked. One can explore ideas that have made their way into chemistry even before their foreseen application in quantum computing, such as rotations to more suitable bases and decompositions of the rank-4 Coulomb interaction tensor,\cite{Huggins2021, lee2020efficient} which is the chief responsible for the size scaling in the Hamiltonian. Other ideas came about taking advantage of the knowledge about the measurement process itself, while also sharing some foundational aspects in tensor algebra.\cite{kandala2017hardware} The proposal by Bravyi et al. allows one to eliminate redundant degrees of freedom that are identified with $\mathbb{Z}_2$ symmetries,\cite{tapering} which has been extended to chemistry applications by also considering molecular point-group symmetry.\cite{point_group_symmetry} Much progress has also been made in searching for the groups of commuting observables, which share the same eigenbasis and can thus be measured upon implementation of a single circuit instance.\cite{pranav, izmaylov1, izmaylov2, izmaylov3} A recent detailed study discussing the role of measurements in chemistry, with emphasis on the VQE algorithm, provides measurement estimates for some common molecules.\cite{gonthier2020identifying}

Briefly touching on the topic of quantum hardware, we mention that a few architectural proposals have matured enough to the point where they are routinely employed. The leading technologies are undoubtedly those based on superconductors and trapped ions, with more recent photonic processors.\cite{Arrazola2021} A ground-breaking experiment of quantum computing in the chemistry realm was the demonstration of the VQE algorithm using a photonic chip,\cite{vqe} and the status of chemistry as one of the most promising outcomes of quantum computing gained credence with applications on superconduting hardware.\cite{mccaskey2019quantum} Since then, the scale and complexity of the simulations have certainly increased by leaps and bounds. Some notable examples showcasing superconducting quantum processors are the execution of VQE with the hardware-efficient ansatz for H$_2$, LiH, and BeH$_2$, \cite{kandala2017hardware} and quantum simulation of the Hartree-Fock states of H$_n$, $n$=6, 8, 10, 12 as well the potential energy curve in the cis-trans isomerization of diazene.\cite{google_hf} Molecules as large as water have been reported simulated using processors build on ion-traps.\cite{ionq} Benchmarks detailing the accuracy and the demands of different processors for chemical applications have also been reported.\cite{ornl1, ornl2}

In the same way advancement in quantum chemistry on classical hardware can be credited to a great extent to the interplay of theory, hardware, and software, quantum computation for chemistry applications gives indication to follow a similar path. Concrete implementations of hybrid quantum-classical algorithms involve the concerted operation between radically different architectures, which in turn produce and manipulate data in distinct ways, often handled by disparate programming languages. Up to this point, the software landscape in the quantum computing domain with relevance to chemistry has been largely dominated by frameworks written in Python, mainly due to user-friendliness and ease of prototyping and potential rapid development, among which we can name OpenFermion,\cite{openfermion} Qiskit,\cite{Qiskit} Pennylane,\cite{pennylane} and TEQUILA.\cite{tequila} Another reason for the adoption of Python in many frameworks with relevance in quantum chemistry is that some popular and open-source quantum chemical suites also provide a Python front-end with a comprehensive set of APIs, with two noteworthy examples being Psi4\cite{psi4} and PySCF.\cite{pyscf}

Despite these appealing properties, in practice the use of Python comes at the expense of unsatisfactory performance, and by virtue of being a high-level language (high-level here means it is far removed from the actual language level at which the hardware operates), it does not exhibit the characteristics required of software in order to be robust and performant in this context. An alternative that addresses these demands is the XACC C\texttt{++} framework,\cite{xacc1, xacc2} which offers users with many of the state-of-the-art quantum algorithms for quantum chemistry,\cite{claudino2021backendagnostic} all the while providing the same interface for virtual execution as well as actual quantum hardware from several vendors. Moreover, because the orbitals fed into simulations such as discussed here are obtained from HF, which can be carried out in polynomial time, one may expect that in the fault-tolerant regime the HF step will itself need to be highly efficient, demanding quantum chemical codes developed in low level languages, such as those in Fortran and C\texttt{++}.\cite{aces, nwchem, GAMESS}

\section{Conclusions and outlook}

Quantum computation is in its infancy, and so are its applications touching on quantum chemistry and many-body physics. This means that what has been presented here is nothing but a tiny patch of a still very much uncharted territory, and there is still much to be discovered and experimented with. One of the implications of this statement, and this can be observed on a daily basis, is that the field is evolving at a very fast pace. This may make it difficult for newcomers to situate themselves and climb the associated learning curve, being able to catch up with the literature in order to appreciate and eventually be able to contribute in the forefront of the developments in quantum computing. As far as the quantum chemistry community is concerned, this review aims at helping bridge this gap by providing a gentle introduction to the topic.

Due to the rapid developments in the field, it is likely that some of the discussion concerning the current landscape and what is perceived as the state-of-the-art will eventually be seen as dated or inadequate, even more so with the fault-tolerant regime inching closer. However, this review should remain a valuable contribution to introduce chemists, or anyone whose interest is to get acquainted to the basics of gate-based quantum computing with a similar background. Other architectural and computational paradigms exist, and a good example would be those grounded on the adiabatic quantum computing,\cite{adiabatic1, adiabatic2, annealing1, adiabatic_annealing} which has also been employed for molecular electronic structure simulations,\cite{chem_annealing1, chem_annealing2, chem_annealing3, chem_annealing4} but without the same momentum or appeal garnered by logic quantum computers. This further advances the claim that the introductory sections of this review are likely to remain relevant in the near future insofar as one seeks this type of computation.

\section{Funding Information}

This work was supported by the “Embedding Quantum Computing into Many-body Frameworks for Strongly Correlated Molecular and Materials Systems” project, which is funded by the U.S. Department of Energy (DOE), Office of Science, Office of Basic Energy Sciences, the Division of Chemical Sciences, Geosciences, and Biosciences.

\section{Acknowledgements}

The author is grateful to Varun Rishi, Travis Humble, and Alex McCaskey for feedback and/or discussions that helped shape this manuscript.

\bibliography{refs}

\end{document}